\documentclass[preprint,pre,showpacs,floatfix]{revtex4}
\usepackage{graphicx}
\usepackage{dcolumn}
\usepackage{amsmath}
\usepackage{latexsym}
\usepackage{amssymb}
\usepackage{epsfig}
\usepackage{subfigure} 

\newcommand{\be}{\begin{equation}}
\newcommand{\ee}{\end{equation}}

\begin{document}

\title{The Isotropic-Nematic Interface with an Oblique Anchoring Condition}

\author{ S.M. Kamil, A. K. Bhattacharjee, R. Adhikari and Gautam I. Menon}
\affiliation{
The Institute of Mathematical Sciences, CIT Campus, 
Taramani, Chennai 600 113, India
}
\date{\today}
\begin{abstract}
We present numerical and analytic results for uniaxial and biaxial order at the 
isotropic-nematic interface within Ginzburg-Landau-de 
Gennes theory. We study the case
where an oblique anchoring condition is imposed asymptotically on the 
nematic side of the interface, reproducing results of previous work when
this condition reduces to  planar or homoeotropic anchoring. We construct
physically motivated and computationally flexible 
variational profiles for uniaxial and biaxial  order, comparing our variational results
to  numerical results obtained
from a  minimization of the Ginzburg-Landau-de Gennes free energy. While 
spatial variations of the scalar uniaxial and biaxial order parameters
are  confined to the neighbourhood of the interface, nematic elasticity
requires that the director orientation interpolate linearly between either planar or homoeotropic anchoring at the
location of the interface and the imposed boundary condition at infinity. The selection of planar or homoeotropic
anchoring at the interface is  governed by the sign of the Ginzburg-Landau-de Gennes elastic
coefficient $L_2$. Our variational calculations are in close
agreement with our numerics and agree qualitatively with results from density functional
theory and molecular simulations.

\end{abstract} 
\maketitle

\section{Introduction}

Nematic liquid crystals,  typically formed in suspensions of rod-like molecules  whose  aspect ratio 
deviates sufficiently from unity, exhibit orientational order in the absence of translational 
order\cite{degenpro,pcl,kleman}.  Such order is 
quantified through a traceless, symmetric tensor
$Q_{\alpha\beta}$ defined at every point in space\cite{degenpro,gralondej}.
In the nematic phase, the order parameter is
\be
Q_{\alpha\beta} = \frac{3 S}{2} \left ( n_\alpha n_\beta - \frac{1}{3}\delta_{\alpha \beta} \right )
+ \frac{T}{2}  \left ( l_\alpha l_\beta -  m_\alpha m_\beta \right )
\label{definition}
\ee
where the director {\bf n} is defined as the normalized eigenvector corresponding to
the largest eigenvalue of ${\bf Q}$, the subdirector {\bf l} is associated with the sub-leading eigenvalue, and
their mutual  normal {\bf m} is obtained from  {\bf n} $\times$ {\bf l}. The quantities $S$ and $T$ represent the strength of uniaxial 
 and biaxial ordering: $|S| \neq 0$, $T=0$ is the uniaxial nematic whereas $S,T \neq 0$  with $T < 3S$ defines the
biaxial case\cite{degenpro}.

The description of the early stages of phase-ordering upon quenches from the isotropic
phase, the properties of nematic droplets within the isotropic phase and the  structure of the
isotropic-nematic interface are all problems which require that nematic and isotropic phases be treated within the same 
framework. The inhomogeneous order parameter configurations obtained in these cases 
are weighted by the Ginzburg-Landau-de Gennes 
(GLdG) free energy,  obtained  {\em via}  a gradient
expansion in ${\bf Q}$ in which only low-order symmetry allowed terms are retained\cite{degenpro,degennes}.
The simplest of the problems  above is that of the structure of the infinite, flat isotropic-nematic interface,
studied initially by de Gennes\cite{degennes}.  

Nematic ordering is strongly  influenced by confining walls and  surfaces, which impose a preferred 
orientation or ``anchoring condition'' on the nematic state. Such a preferred orientation yields an anchoring angle, defined 
as the angle made by the  director in the immediate neighbourhood of the surface with the surface normal. 
Anchoring normal  to the surface is termed as homoeotropic,  whereas  anchoring in the plane of the surface is 
termed as planar. The general case is that of oblique anchoring. 

As is the case with surfaces, the interface between a nematic and its isotropic phase can also favour a particular
anchoring.  The problem of interface structure for  the nematic is particularly interesting since it  illustrates  how the structure
in the  interfacial region can differ substantially from structure in the bulk. It is known, for example,  that a region proximate to the
interface can exhibit biaxiality within the LGdG theory, even if the stable nematic phase is pure uniaxial\cite{popasluckwh}, provided  
planar anchoring is enforced. Such biaxiality is absent if the anchoring is homoeotropic\cite{degennes}. These two limits,
of homoeotropic and planar anchoring, lead to interface profiles of $S$ and $T$ which vary only in the
vicinity of the interface, as well as orientations which are uniform across the interface\cite{degennes}.

Can oblique anchoring be stabilized, within GLdG theory, at the interface between a bulk uniaxial nematic and its isotropic phase? 
Suppose we introduce boundary conditions that impose a specified  oblique 
orientation deep into the nematic phase, where the magnitude of the order parameter  is saturated. 
The question, then, is whether such an imposed orientation is relaxed to a preferred value in the vicinity of the 
interface.  The difficulties with this problem stem from the fact that  changes in the local frame orientation on the 
nematic side of the interface come with an elastic cost arising out of nematic elasticity. This is an effect sensitive, in principle,
to system dimensions, since gradients can be smoothed out by allowing the changes to occur over the system size. 
While this cost can be reduced by suppressing  the order parameter amplitudes in regions where order parameter
phases vary strongly, the precise way in which this might happen, if at all, is an open question.

Popa-Nita, Sluckin and Wheeler  (PSW)\cite{popasluckwh} studied  this problem numerically within a GLdG  
approach, using a set of variables $\eta_s$ and $\mu_s$ introduced in Ref.~\cite{sensullivan}. These variables 
are  combinations of the variables $S$, $T$ and $\theta$ used in this paper.   Although the
focus of their study was the emergence of biaxiality at the interface with a planar anchoring condition, PSW remarked
that if the asymptotic orientation of the director in the nematic phase was set to any value other than $90^{\circ}$ (planar anchoring) 
or $0$ (homoeotropic anchoring) for large $z$, then $\eta_s$ and $\mu_s$ approached this value with non-zero slope. PSW thus 
concluded that there could be no stable anchoring if the orientation of the director in the nematic phase was neither planar 
nor homoeotropic, but oblique. The precise nature of the resulting state obtained upon applying an oblique anchoring condition
was not addressed by PSW\cite{popasluck,popasluckwh}.

Density functional calculations on hard-rod systems using Onsager's theory applied to the free isotropic-nematic interface
indicate that the minimum surface free energy is obtained when the rods lie parallel to the isotropic-nematic interface, the 
case of planar anchoring\cite{allen2000a,allen2000b}. Molecular simulations of a system of hard ellipsoids, in which 
an anchoring energy fixes the director orientation in the nematic phase at a variety of angles, indicate that the isotropic-nematic
interface favours planar anchoring. These simulations, and a mean-field calculation based on the
Onsager functional, find that the angle profile is approximately linear as one moves away from the boundary condition
imposed by the wall at one end of the simulation box\cite{schilling,velasco}. These results, in particular 
concerning the stability of planar anchoring, are consistent with those from other  
treatments\cite{bateszannoni,chen1,chen2,albarwani,vinkschilling}. However, several other papers indicate
specific regimes in which homoeotropic or oblique anchoring may be stable. 
Moore and McMullen\cite{mooremcmullen} numerically evaluate the inhomogeneous grand potential within a specific
approximation scheme finding that  planar anchoring is preferred at the interface for long spherocylinders, but 
oblique or homoeotropic anchoring may be an energetically favourable alternative for smaller aspect ratios. Holyst and
Poniewierski study such hard spherocylinders in the Onsager limit, noting that  oblique anchoring  is favoured over a considerable
range of aspect ratios\cite{holyst}. Finally, experiments provide evidence for both oblique\cite{faetti} and planar anchoring\cite{langevin}, 
with electrostatic effects possibly favouring oblique anchoring.

\begin{figure}
\includegraphics[width = 5 in]{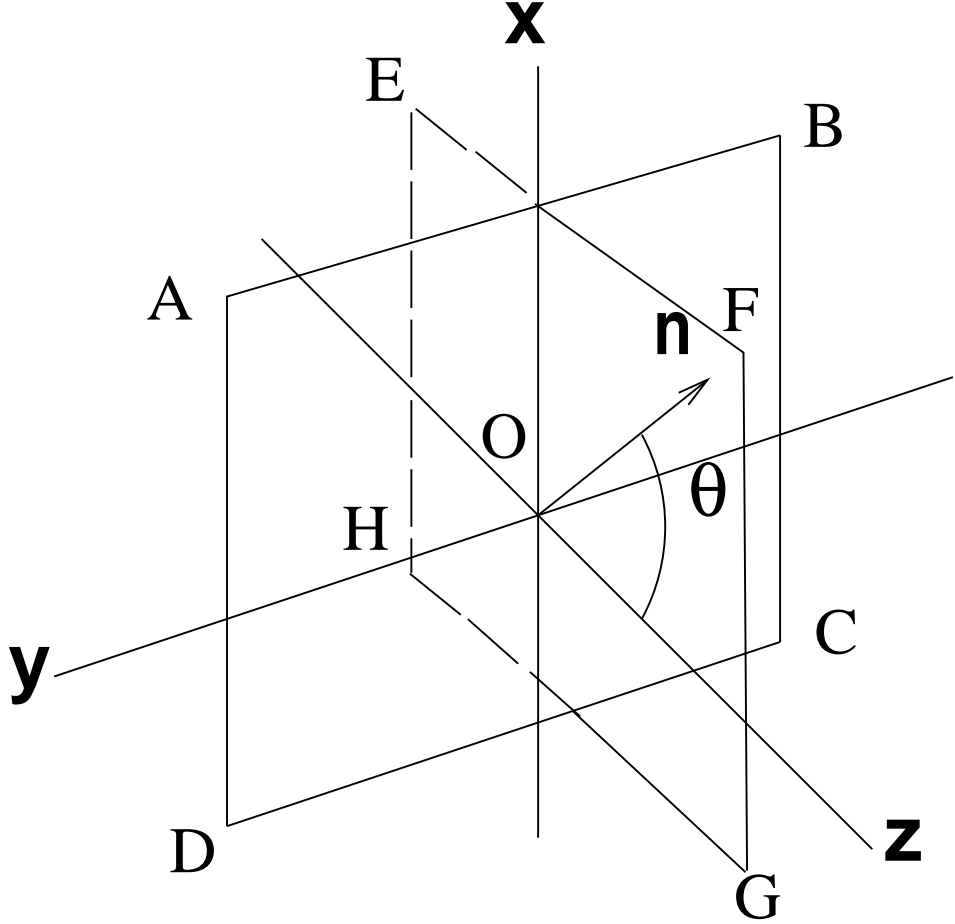}
\caption{The interface geometry and the coordinate system used in this paper. Note that the nematic director
makes an angle $\theta$ with respect to the $z-$ axis direction. This angle is fixed at infinity to $\theta = \theta_e$ . It can be
chosen to  vary between $\theta_e = 0$ (homoeotropic anchoring at infinity) and $\theta_e = 90^\circ$ (planar anchoring at infinity). The isotropic 
phase is favoured, through boundary conditions, as $z \rightarrow -\infty$, whereas the nematic phase is favoured for
$z \rightarrow \infty$. The plane of the interface is the $x-y$ plane, shown by ABCD in the figure, whereas the director is
confined to the EFGH plane as shown. The origin is denoted by O.}
\label{labframe}
\end{figure}

This paper  studies the isotropic-nematic interface within GLdG theory in the case where an  oblique anchoring condition is imposed on the 
nematic state far from the location of the interface.  For a flat interface, the components of ${\bf Q}$ can  depend only on the 
coordinate perpendicular  to the interface. We assume that this coordinate is aligned along the $z$ axis, as shown in  Fig.~\ref{labframe},
which defines the geometry we work with in this paper.
We work at phase coexistence, imposing  boundary conditions fixing  the isotropic phase at $z$ = $-\infty$ and the nematic phase at $z$ = $\infty$. 
The components
of ${\bf Q}$ as $z \rightarrow \infty$  are chosen so that $S$ is fixed to its value at coexistence $S_c$, while the
axis of the nematic is aligned along a specified (oblique) direction.  
The coexisting states must be separated by an interface in which order parameters 
rise from zero on the isotropic side of the interface to saturated, non-zero values on the nematic
side. Since the two free energy minimum states are degenerate in the
bulk,  the position of the interface is arbitrary and can be fixed, for concreteness, at $z = 0$ in the infinite system.  However, there are 
subtleties.  Provided all components of ${\bf Q}$  vary substantially only  in the
neighbourhood of the interface, the  interface can be located through several, largely equivalent criteria. However, if 
variations of ${\bf Q}$  are not confined to a region proximate to the interface but depend on the system size
$L$ irrespective of how large $L$ is, the very isolation of an interface from the bulk  is  ill-defined. As indicated
earlier, it is this situation which obtains in the case of oblique anchoring and the
$L \rightarrow \infty$ limit must be taken with care.

The central results of this paper are the following: A numerical minimization of the GLdG free energy
which imposes a specific  oblique anchoring 
condition on the system deep into the nematic while fixing the interface location at the origin
shows that  the  elements of ${\bf Q}$ vary with space even far away from the interface, albeit slowly. Only in the
limit of homoeotropic or planar anchoring is the variation of ${\bf Q}$ confined to a finite region. This variation in
the case of oblique anchoring can, however,
be split into hydrodynamic and non-hydrodynamic components. Generically, the variation of the non-hydrodynamic
components, such as the magnitudes of $S$ and $T$, are confined to a finite region, independent of the system size $L$,
if $L$ is large enough. However, the orientation of the nematic director varies in space:  if the asymptotic value of the nematic 
order parameter at $L$ represents uniaxial ordering along an oblique axis,  the director orientation  interpolates
linearly between either a ${90^\circ}$ value preferred at the location of the interface (planar
anchoring) or a ${0^\circ}$ value (homoeotropic anchoring), and the value imposed by the boundary  condition at $L$. 
Whether planar or a homeoetropic anchoring is preferred at the interface depends on the sign of the second of
the elastic coefficients in the GLdG expansion, the $L_2$ term, as initially shown by de Gennes\cite{degennes}. 

Our results are consistent with the qualitative observations of PWS, but provide
a detailed quantitative analysis in the case of oblique anchoring. We scale
angle profiles computed for different values of the system size $L$  onto a universal curve, indicating a linear
profile. In the limit that $L \rightarrow \infty$, the slope with which the phase
varies vanishes as $1/L$, so that the total energy cost for elastic distortions of the nematic field $\sim \int  (\nabla \theta)^2 dz
\sim L (\Delta \theta)^2/L^2 \sim 1/L$, thus vanishing in the thermodynamic limit. Thus, the isotropic-nematic interface
with an oblique anchoring constraint imposed on the nematic side can be regarded as being marginally stable, as opposed to
unstable, provided the thermodynamic limit is taken with care.
We demonstrate that suitably chosen, flexible variational
choices for the uniaxial and biaxial profiles can capture the variation of components of the {\bf Q} tensor as a function of
space. These variational profiles are obtained by generalizing results from a calculation of biaxial and
uniaxial order parameter profiles in the planar case. These profiles are benchmarked against numerical calculations. 

The outline of this paper is the following. In Section \ref{sectionGL}, we briefly review aspects of the
Landau-Ginzburg-de Gennes transition which will be required in our analysis and obtain the
equations representing the variational minimum of the GLdG free energy, in a basis adapted to the symmetry
of the problem. Section~\ref{sectioninter}  describes solutions to these equations, as appropriate to
the cases of planar and homoeotropic anchoring. The classic $\tanh$ profile obtained by de Gennes
is an exact representation of the interface in the limit of homoeotropic anchoring  as well as when the
 $L_2$ elastic constant vanishes, in which case the interface is stable for any anchoring condition. In Section
 ~\ref{sectionnumer} we present our numerical approach to the problem of interface structure, showing how 
 numerically exact profiles for the variation of $S$, $T$ and $\theta$ can be obtained within the framework of
 a minimization of the full GLdG free energy, subject only to the condition that an interface is forced into the system.
 
 In Section~\ref{sectionvar}, we describe our variational approach to this problem, motivating the choice of
 a three-parameter variational ansatz inspired by the approximate solution due to Popa-Nita, Sluckin and Wheeler.
 We show that this variational ansatz captures the features of the solution in both the extreme cases of
 planar and homoeotropic anchoring, and is flexible enough to describe the intermediate regime as well. 
 In Section~\ref{sectionvarnum}, we describe our methods of minimization for
 the variational problem and  our results for $L_2 >0$ and $L_2 < 0$.  We describe how our numerical and 
 variational calculations can be used to provide an accurate picture of the interface with an oblique 
 anchoring condition{ In Section~\ref{sectionasymp} we present asymptotic results for the variation of $S$, $T$ and
 $\theta$ close to the bulk nematic state. Section~\ref{sectionconclusion} contains  our conclusions.
 
\section{The Ginzburg-Landau-de Gennes Approach to the Isotropic-Nematic Transition}
\label{sectionGL}
The Ginzburg-Landau-de Gennes free energy functional $F = F_h + F_{el}$ \cite{degennes} is obtained from a local expansion
in powers of rotationally invariant combinations of the order parameter ${\bf Q}({\bf x},t)$,
\begin{equation}
\label{localfrener}
\mathcal{F}_{h}[{\bf Q}] = \frac{1}{2}ATr{\bf Q}^{2}  +
\frac{1}{3}BTr{\bf Q}^{3} + \frac{1}{4}C(Tr{\bf Q}^{2})^{2} +
E^{\prime}(Tr{\bf Q}^{3})^{2} \ldots,
\end{equation}
The restriction to the terms shown above are sufficient to yield  a first-order transition between isotropic
and nematic phases as well as a  stable biaxial phase, obtained when $E^\prime \neq 0$\cite{gralondej}. 

To this  local free energy, non-local terms arising from rotationally invariant combinations of gradients of
the order parameter must be added. The choice of the following two lowest-order gradient terms is common\cite{degennes,popasluck,popasluckwh}:
\begin{equation}
\mathcal{F}_{el}[{\bf \partial Q}] = \frac{1}{2}L_{1}(\partial_{\alpha}
Q_{\beta\gamma})(\partial_{\alpha}Q_{\beta\gamma}) +
\frac{1}{2}L_{2}(\partial_{\alpha}Q_{\alpha\beta})(\partial_{
\gamma}Q_{\beta\gamma}),
\end{equation}
where $\alpha, \beta, \gamma$ denote the Cartesian directions in the local
frame, and $L_{1}$ and $L_{2}$ represent the elastic cost for distortions in {\bf Q}\cite{gralondej}. 
The fact that there are only two terms which appear to this order implies that only two of the three Frank
constants are independent. The limit in which $L_2 = 0$, or of zero elastic anisotropy corresponds to the
case in which all Frank constants are equal. The relationship between $L_1$ and $L_2$ and the
Frank constants $K_1,K_2$ and $K_3$ are the following: $K_1 = K_2 = 9/4(2L_1 + L_2)S^2$
and $K_3 = 9/2L_1 S^2$\cite{degenpro,gralondej}. Note that $\kappa = L_2/L_1$ negative is allowed, although $\kappa < 1.5$ must
be satisfied to ensure positivity of the elastic constants.

In the free energy density of Eq.~ \ref{localfrener}, $A = A_{0}(1 - T/T^{*})$, 
where $T^{*}$ denotes the supercooling transition temperature. From the inequality
$\frac{1}{6}(Tr{\bf Q}^{2})^{3} \geq (Tr {\bf Q}^{3})^{2}$, higher powers of $Tr{\bf Q}^{3}$ 
can be excluded for the description of the uniaxial phase. Thus the uniaxial case is 
described by $E^{\prime}$ = 0 whereas $E^{\prime} \neq 0$ for the biaxial phase. We will
assume that $E^{\prime}$ = 0, thus ensuring that the stable ordered phase is the uniaxial
nematic. For nematic rod-like molecules $B<0$ whereas for disc-like molecules, $B>0$; for concreteness,
we will assume $B < 0$ here. The 
quantity C  must be positive  to ensure 
stability and boundedness of the free energy in both the isotropic and nematic phases.

The first order isotropic to uniaxial nematic transition at the critical value $S = S_{c}$ 
is thus obtained from,
\begin{eqnarray}
A &=& \frac{3}{4}CS_c^{2}  \\
B &=& -\frac{9}{2}CS_c.
\end{eqnarray}
We choose
$B = -0.5, C= 2.67$ and $A = B^2/27C$, thus enforcing phase coexistence between an isotropic and
uniaxial nematic phase \cite{gralondej}. 

The interface is taken to be flat and infinitely extended in the $x-y$ plane. 
The spatial variation of the order parameter only occurs along the $z$ direction\cite{sensullivan}. 
We scale $Q_{\alpha \beta} \rightarrow  Q_{\alpha \beta}/S_c$ where $S_c = -\frac{2B}{9C}$,
$\mathcal{F}  \rightarrow  \frac{16}{9CS_c^4}\mathcal{F}$, and measure lengths in units of
$l_c = \sqrt{54 C (L_1 + 2L_2/3)/B^2}$. 

\subsection{The Ginzburg-Landau-de Gennes Equations}
\label{sectionVar}
The director ${\bf n}$, sub-director ${\bf l}$ and their joint normal ${\bf m}$ together define 
a frame. We define  $z$ as the direction perpendicular
to the interface.  The fixed orientation of the nematic axis at $z \rightarrow \infty$
can be used to define a plane, the $xz$ plane.  From symmetry, and following the
arguments of Sen and Sullivan, the nematic director 
must always remain in this plane\cite{sensullivan}. Thus, the  spatial dependence of the frame 
orientation can only come from the variation of a single  tilt angle $\theta$, which is measured
between the $z$ axis and ${\bf n}$, as shown in Fig.~\ref{labframe}.

Since we assume a flat interface,  the components of $\bf{Q}$ are functions only of $z$. 
The tensor $\bf{Q}$ n the local frame defined by the principal axes, is diagonal and given  by 
\begin{equation}
{\bf Q} = \left(\begin{array}{ccc} -(S +T)/2& 0 & 0 \\ 0 & -(S - T)/2 & 0\\ 0 & 0 & S \end{array}\right)
\end{equation}
Transforming to the space-fixed frame (the laboratory frame), by rotation through the appropriate
angle $\theta$ yields
\begin{equation}
\bf{Q}_\theta = {\small{}
\left(\begin{array}{ccc} \cos{\theta}& 0 & \sin{\theta} \\ 0 & 1 & 0\\ -\sin{\theta} & 0 & \cos{\theta} \end{array}\right)
\left(\begin{array}{ccc} -(S +T)/2& 0 & 0 \\ 0 & -(S - T)/2 & 0\\ 0 & 0 & S \end{array}\right)
\left(\begin{array}{ccc} \cos{\theta}& 0 & -\sin{\theta} \\ 0 & 1 & 0\\ \sin{\theta} & 0 & \cos{\theta} \end{array}\right)}.
\end{equation}
Thus,   $\bf{Q}_\theta$ takes the form
\begin{equation}
{\bf Q_{\theta}} = \left(\begin{array}{ccc} -\frac{1}{2}(S +T)\cos^2{\theta} + S \sin^2{\theta}& 0 & -\frac{1}{4}(3S+T)\sin{2\theta} \\ 0 & -(S - T)/2 & 0\\ -\frac{1}{4}(3S+T)\sin{2\theta} & 0 & -\frac{1}{2}(S +T)\sin^2{\theta} + S \cos^2{\theta} \end{array}\right).
\end{equation}
Inserting this tensor form into the elastic free energy $\mathcal{F}_{el}[{\bf Q}]$ yields the elastic contribution to the free energy
\begin{eqnarray}
F_{g\theta} & = &\frac {(12 + 5 k + 3 k\cos(2\theta)) {\partial_z S}^2 + 4 k \sin^2(\theta) {\partial_z S} {\partial_z T} + 2\left (2 + k\sin^2(\theta) \right) {\partial_z T}^2} {8 (3 + 2 k)} \nonumber \\ 
 & &- \frac {2 k\sin(2\theta) (3 S + T)\left ({\partial_z S} - {\partial_z T} \right) {\partial_z\theta}} {8 (3 + 2 k)} 
+ \frac {(2 + k) (3 S + T)^2{\partial_z\theta}^2} {4 (3 + 2 k)},
\label{elastic_free}
\end{eqnarray}
Note that this contribution must vanish if $S,T$ and $\theta$ are uniform.

The bulk free energy contribution $\mathcal{F}_{h}[{\bf Q}]$ is unchanged, as a consequence of
the fact that the Landau term is constructed from rotationally invariant terms in the order parameter.
It then takes the form
\begin{equation}
\mathcal{F}_{h}[{\bf Q}] =\frac{1}{3}(3{S}^2 + {T}^2) - 2({S}^3 - {S}{T}^2) + \frac{1}{9}(9{S}^4 + 6{S}^2{T}^2 + {T}^4).
\label{elastic_landau}
\end{equation}
The Euler-Lagrange equations minimizing the full free energy $F$, are obtained from
\begin{equation}
  -(\frac{\partial F_{g\theta}}{\partial{\theta}}) +\frac{d}{dz}(\frac{\partial F_{g\theta}}{\partial \dot{\theta}}) = 0,
\end{equation}
where $\dot{\theta} = d\theta/dz$.
This yields
\begin{eqnarray}
4 (2 + k)\left (3 S' + T' \right) \theta'- k \sin(2\theta) (S''- T'')  
+  2(2 + k) (3 S + T)\theta'' = 0,
\end{eqnarray}
which can further be simplified as 
\begin{eqnarray}
- k \sin(2\theta) (3 S + T) (S''- T'') + (2(2 + k) (3 S + T)^2\theta')' = 0,
\end{eqnarray}
where the primes indicate derivatives with respect to $z$. 

First, note that  for $k = 0$  ({\em i.e.} no elastic anisotropy) the above equation has only the 
solution $\theta' = 0$, implying that $\theta$ is constant. 
A similar situation holds for the special $\theta$ values  $\theta = 0,  90^\circ$,
for which again the only solution has $\theta' = 0$.  Thus, in these special limits, the angle
$\theta$ remains fixed throughout the system. These results are, of course, consistent with the result that planar ($\theta = 90^\circ$) and
homoeotropic ($\theta = 0$) anchoring conditions yield a well-defined
interface. Also, provided elastic anisotropy is absent, one can continue to define a stable
interface for an arbitrary $\theta$, since $\theta$ sticks to its asymptotic value throughout.

Finally, we note that once $S$ and $T$ are saturated, $S' = T' = S'' = T'' = 0$, and thus
$\theta'$ = constant, yielding a linear variation of $\theta$ with $z$.

For completeness, the full set of Euler-Lagrange equations representing the minimization
of the GLdG equations are, in addition to the $\theta$ equation above
\begin{eqnarray}
- \left (\frac { (k\cos(2\theta)+6 +3 k)\left (3S + T \right)} {6 + 4 k} \right)\theta'^2 +
 \left (\frac {k (4 + 3 k + k\cos(2\theta))\sin(\theta)^2} {4\left (6 + 7 k + 2 k^2 \right)} \right) T'' \nonumber \\
+ \left (\frac {96 + 88 k + 19 k^2 + 12 k (2 + k)\cos(2\theta) + k^2\cos(4\theta)} {16\left (6 + 7 k + 2 k^2 \right)} \right) S''
 = 2 S - 6 S^2 + 4 S^3 + 2 T^2 + 4 S\frac {T^2} {3} \nonumber \\ 
\end{eqnarray}

\begin{eqnarray}
\left (\frac {32 + 24 k + 3 k^2 - 4 k (2 + k)\cos(2\theta) + k^2\cos(4\theta)} {16\left (6 + 7 k + 2 k^2 \right)} \right) T'' 
+ \frac {k (4 + 3 k + k\cos(2\theta))\sin(\theta)^2 S''} {4\left (6 + 7 k + 2 k^2 \right)}
\nonumber \\
 + \frac {(k\cos(2\theta)-2-k) (3 S + T)\theta'^2} {6 + 4 k} 
 = \frac {2} {3} T + 4 S\text {} T + \frac {4} {9} T^3 + \frac {4} {3} S^2 T \nonumber \\
\end{eqnarray}

\section{Interface structure for Planar and Homoeotropic Anchoring}
\label{sectioninter}
This section briefly reviews the methodology for the solution of interfacial
structure in the cases of homoeotropic and planar alignment\cite{degennes}. While the
exact solution in the case of homoetropic alignment, as originally
proposed by de Gennes, motivates the canonical $\tanh$ form  for 
the uniaxial order parameter, the more complex situation of planar
anchoring requires the simultaneous solution of equations of motion
for both $S$ and $T$, in addition to the equation for $\theta$\cite{popasluckwh}. We discuss how the Popa-Nita, Sluckin and Wheeler
solution\cite{popasluckwh} of the planar case can be generalized, in a variational
sense, to the more general problem of an oblique anchoring condition. 

\subsection{Homeotropic Alignment}

The equation of motion for homoeotropic boundary conditions is easily obtained
by setting $\theta=0$, in the defining equations above. This immediately
yields,
\begin{eqnarray}
\frac{1}{2}\partial_z^2S  & = & S - 3 S^2 + 2 S^3 +  T^2 + \frac{2ST^2}{3},
\label{parfirst}
\end{eqnarray}
\begin{eqnarray}
\frac{1}{2(3 + 2 k)}\partial_z^2T & = & \frac{1}{3} T + 2 S T + \frac{2 T^3}{9} + \frac{2 S^2 T}{3}.
\label{parsecond}
\end{eqnarray}
It is easy to see that these equations have the solutions
\begin{equation}
S = \frac{1}{2}(1+\tanh(\frac{z}{\sqrt{2}})),\quad T = 0;
\end{equation}
Here the treatment of de Gennes is exact.

\subsection{Planar Alignment}
The case of planar alignment follows from setting $\theta = 90^\circ$ in the
Euler-Lagrange equations. This then yields the following set of coupled
partial differential equations for the $S$ and $T$ order parameters,
\begin{eqnarray}
\frac{(6 + k)}{(3 + 2 k)}\partial_z^2S +\frac{k}{(3 + 2 k)}\partial_z^2 T & = & 4 S - 12 S^2 + 8 S^3 + 4 T^2 + \frac{8ST^2}{3},
\label{ndfirst1}
\end{eqnarray}
\begin{eqnarray}
\frac{k}{(3 + 2 k)}\partial_z^2 S +\frac{(2 + k)}{(3 + 2 k)}\partial_z^2T & = & \frac{4}{3} T + 8 S T + \frac{8 T^3}{9} + \frac{8S^2 T}{3}.
\label{ndsecond2}
\end{eqnarray}

In the zeroth order aproximation we drop terms in $T$ as in the solution of the first equation. This then yields
$S = \frac{S_c}{2}(1 + \tanh(\frac{z}{\sqrt{2}\xi}))$
where $\xi = \sqrt{\frac{1 +k/6}{1+2k/3}} $. Putting this in equation (\ref{ndsecond2}), scaling z again with $\sqrt{2} \xi$
and neglecting the nonlinear term, we get the following equation.
\begin{eqnarray}
\partial^2_z T & = &2\beta (\tanh^2(z) + 8 \tanh(z) + 9 )T \nonumber \\ 
                 & & + \frac{k}{2+k}\tanh(z)(1+ \tanh(z))(1 - \tanh(z)),
\label{ndthird}
\end{eqnarray}
with $\beta = \frac{6+k}{3(2+k)}$.

The PSW approximation now consists of dropping the $\partial^2_z T$ term, yielding the
algebraic equation
\be
2\beta (\tanh^2(z) + 8 \tanh(z) + 9 )T 
                  = - \frac{k}{2+k}\tanh(z)(1+ \tanh(z))(1 - \tanh(z)),
\label{ndfourth}
\ee
which then immediately yields
\be
T  = - \frac{k}{2\beta (2+k)}\frac{\tanh(z)(1+ \tanh(z))(1 - \tanh(z))}{(\tanh^2(z) + 8 \tanh(z) + 9 )}.
\label{ndfourth}
\ee
We have recently suggested an improvement to these results, motivated by our tests of the self-consistency
of the PSW approximations. These tests indicate that the $\partial^2_z T$ term dropped by PWS should be retained for 
a more accurate description of the interface. Our analytic results for this case, expressed as a sum over
hypergeometric functions, agree well with  numerical solutions of the GLdG equations and represent a significant
improvement over the PSW solution, particularly in the case of small $\kappa$.

%%%%%%%%
\section{Numerical Minimization of the Ginzburg-Landau-de Gennes Free Energy for the Interface Problem}
\label{sectionnumer}
Our numerical results for the isotropic-nematic interface with an oblique anchoring condition are obtained from a
direct minimization of the Ginzburg-Landau-de Gennes functional, with boundary conditions which ensure
the presence of the interface as well as impose the required anchoring condition on the $\theta$ field.
Our numerical methodology is the following: Defining a system size $L$, we discretize the one-dimensional (z) 
coordinate into  $N$  points, defining $\delta = L/N$. We use, typically, $N = 1001$. The values of the fields $S$, 
$T$ and $\theta$ at each of these points is
varied so as to minimize the combined integrals of Eq. ~\ref{elastic_free} and Eq. ~\ref{elastic_landau}.

\begin{figure}
\includegraphics[width = 7.5 in]{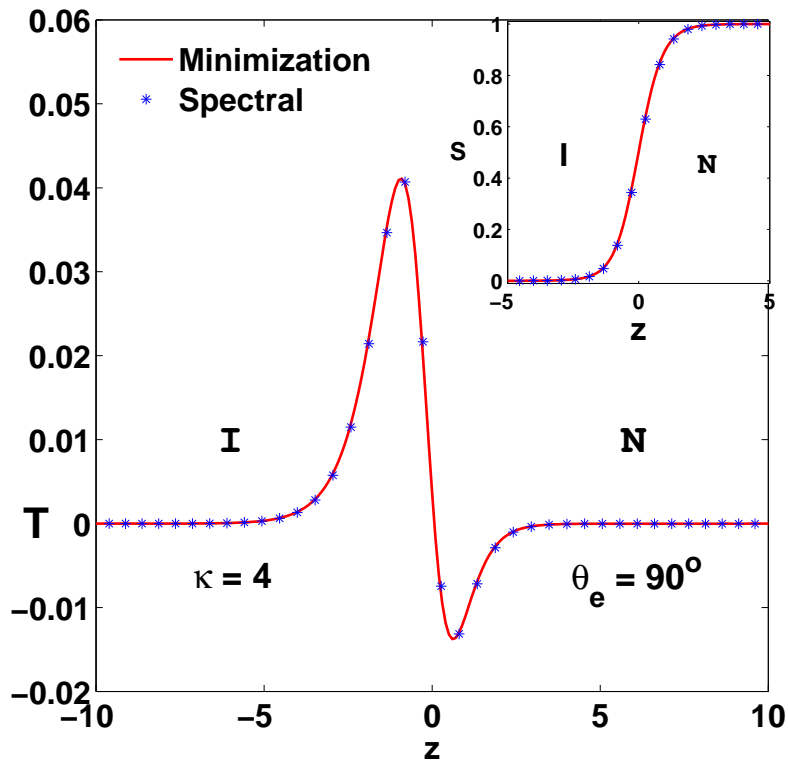}
\caption{ [Color online] Profiles of the biaxial ($T$) (main figure) and uniaxial ($S$) order (inset)
parameter as a function of the coordinate $z$ across the interface, for planar anchoring and
 $\kappa =4$ as  obtained from a  direct numerical minimization of the LGdG functional (solid line).
The results obtained from an  spectral collocation method are shown as points. }
\label{AKTk4angle90}
\end{figure}

To do this, we perform a straightforward evaluation of the integral using the trapezoidal rule, replacing
derivative terms in the integrand by the finite difference approximants. Thus, the gradient term
$dS/dz \simeq \left [S(i + 1) - S(i)\right ]/\delta $. We have also used a variable discretization in some of our
calculations, to assess the accuracy of our results, sampling with closely spaced points in the vicinity of the
interface where the variation of $S$ and $T$ is largest. We impose boundary conditions on $S$, $T$ and $\theta$,
by fixing the values at the two extreme boundaries to their values in the isotropic ($S=0,T=0$) limit, with $\theta$
arbitrary, and in the nematic limit ($S=1,T=0,\theta = \theta_e$). 

The location of the interface is  fixed at the centre, by imposing $S = 1/2$ at the central site. In principle, in a 
system of finite size $L$, our methods yield a constrained minimum for the following reason: The elastic
energy on the nematic side is minimized by allowing the nematic region to expand as far as possible, effectively
forcing the interface to invade the isotropic side. However, as discussed above, in the thermodynamic limit 
$L \rightarrow \infty$, this elastic energy cost reduces as $1/L$, vanishing in the thermodynamic limit where a
stable interface is obtained.  Alternatively, one can think of this  in terms of adding a localized pinning 
potential with  strength vanishing as $L \rightarrow \infty$, which serves only to stabilize the location of the interface. 

This relatively simple approach yields results of very high quality, as we have checked by a direct comparison to 
exact results for the planar  anchoring case as well as to numerical calculations using spectral methods in the 
case of planar anchoring. We have used the minimization routines (NMinimize) in Mathematica to find the stationary values of 
$S$,$T$ and $\theta$ which minimize the free energy subject to the applied boundary conditions. This
routine selects the most appropriate methodology from a variety of minimization techniques available,
iterating till an accuracy between successive iterations of 1 part in 10$^8$ is obtained.

As a test of the quality of the minimization methodology which will be used in this paper, 
we show in Fig.~\ref{AKTk4angle90},  profiles of the biaxial ($T$) (main figure) and uniaxial ($S$) order (inset)
parameter as a function of the coordinate $z$ across the interface, as computed by the numerical spectral methodology
of Ref.~\cite{kamil-2009}
and the minimization technique described above,  for the case of planar anchoring {\it i.e.}
$\theta_e= 90^\circ$, with  $\kappa =4$. Results obtained from  the numerical minimization of the LGdG functional 
are shown as the solid line whereas results from the  spectral collocation scheme of Ref.~\cite{kamil-2009}
are shown as points. These coincide to high accuracy.

\section{Variational Method}
\label{sectionvar}
Clearly, the solution of the full set of equations for $S$, $T$ and $\theta$ given above is a formidable problem.
Our approach to this problem therefore proceeds through the construction of simple,
physically motivated variational choices for $\theta(z)$, $S(z)$ and $T(z)$. This choice is made 
keeping in mind that requirement that the results should be consistent with  computations in the simpler
$\theta =0,90^\circ$ limits, where the angular variation is absent and the de Gennes solution and the PSW solution
are obtained, respectively.

Our approach begins by assuming a profile of the form 
\begin{equation}
S = \frac{1}{2}(1+\tanh(a z))\quad \text{and} \quad T = -b \tanh(c z)\frac{(1 + \tanh(c z))(1 - \tanh(c z))}{\tanh^2(c z) + 8 \tanh( c z) + 9}.
\end{equation}
together with  the assumption that the theta variation can be fitted to a simply parametrizable function. We have examined
a variety of such functions for the case of planar anchoring, including (a) $\theta = 90^\circ - 2\frac{\psi}{L}z$ for $z > 0, 90^\circ$ for $z < 0$,
(b) $\theta = 90^\circ - \frac{\psi}{2}(1+\tanh(a_1 z))$ which implies that
at $z =\infty$,  $\theta = 90^\circ -\psi$ and at $z = -\infty$, $\theta = 90^\circ$, (c) $\theta = 90^\circ - \frac{\psi}{2}(1+\tanh(a_1 z))$ which implies that
at $z =\infty$,  $\theta = 90^\circ -\psi$ and at $z = -\infty$, $\theta = 90^\circ$,
(d) $\theta = \frac{\psi}{2}(1+\tanh(a_1 z))$
(e)  $\theta =  \frac{\psi}{2 L}z + \psi/2$ and (f) $\theta = p + \frac{\psi}{2}(1+\tanh(a_1 z))$.
 
Our best results are obtained with  the variational form
\be
\theta = p + \psi \frac{z}{L}
\ee subject to a constraint $p + \psi = \theta_L$ where $\theta_L$ is the value of angle
at $L$, the system size. It will be our intention to take the $L \rightarrow \infty$ limit later.

Note that the choice
$p = 90^\circ, \psi = 0, a= 1, b = \frac{k}{2\beta (2+k)}$ recovers the profile of PSW for the planar case.
The parameter values $\psi = 0, b=0$ generate the de Gennes solution. Thus, the two extreme
limits of the variation of the anchoring angle can be obtained with the appropriate choice of
parameter values in the variational form chosen above. These can be simply generalized to the case of
homoeotropic anchoring. 

\section{Numerical Methodology for the Variational Solution}
\label{sectionvarnum}
These variational {\em ans\"atze} for $S$ and $T$ are inserted into the form for the free energy, which is
then minimized with respect to the parameters $a,b,c,$ and $p$. This minimization is carried out
using Mathematica. We use the "Nelder-Mead" method for the minimization of a function of $n$
variables. This is a direct search method which uses  an initial choice of $n+1$ vectors  which form  
the vertices of a polytope in $n-$dimensions and a methodology for changing the vertices of
this polytope iteratively. The process is assumed to have converged if the difference 
between  the best function values in the new and old polytope,  as well as the distance between the new 
best point and the old  best point, are less than preset values,  typically of the order of $10^{-10}$. 

To eliminate problems arising from an incorrect choice of initial values, we have computed the minima for
about 100 separate initial conditions and chosen the parameter values corresponding to the least
value of the free energy  from these. Our results for the minimization have been crosschecked using the  
differential evolution method, a  simple stochastic function minimizer.

\subsection{Results from the Numerical and Variational Minimization: $\kappa > 0$}
\label{sectionresults}
In Fig.~\ref{jointcollapsT}, we show profiles of the biaxial ($T$) and uniaxial ($S$) order
parameter as a function of the coordinate $z$ across the interface. These are
computed by direct numerical minimization of the LGdG functional, {\em via} the methodology
described in the previous section. We allowed $\theta$ on the isotropic side to vary,
finding that the free energy minimum was obtained when $\theta$ was stuck to the
value it attained at the location of the interface. This value is somewhat smaller than $90^\circ$
for small system sizes but asymptotes to this value as $L$ goes to infinity.

\begin{figure}
\includegraphics[width = 7.5 in]{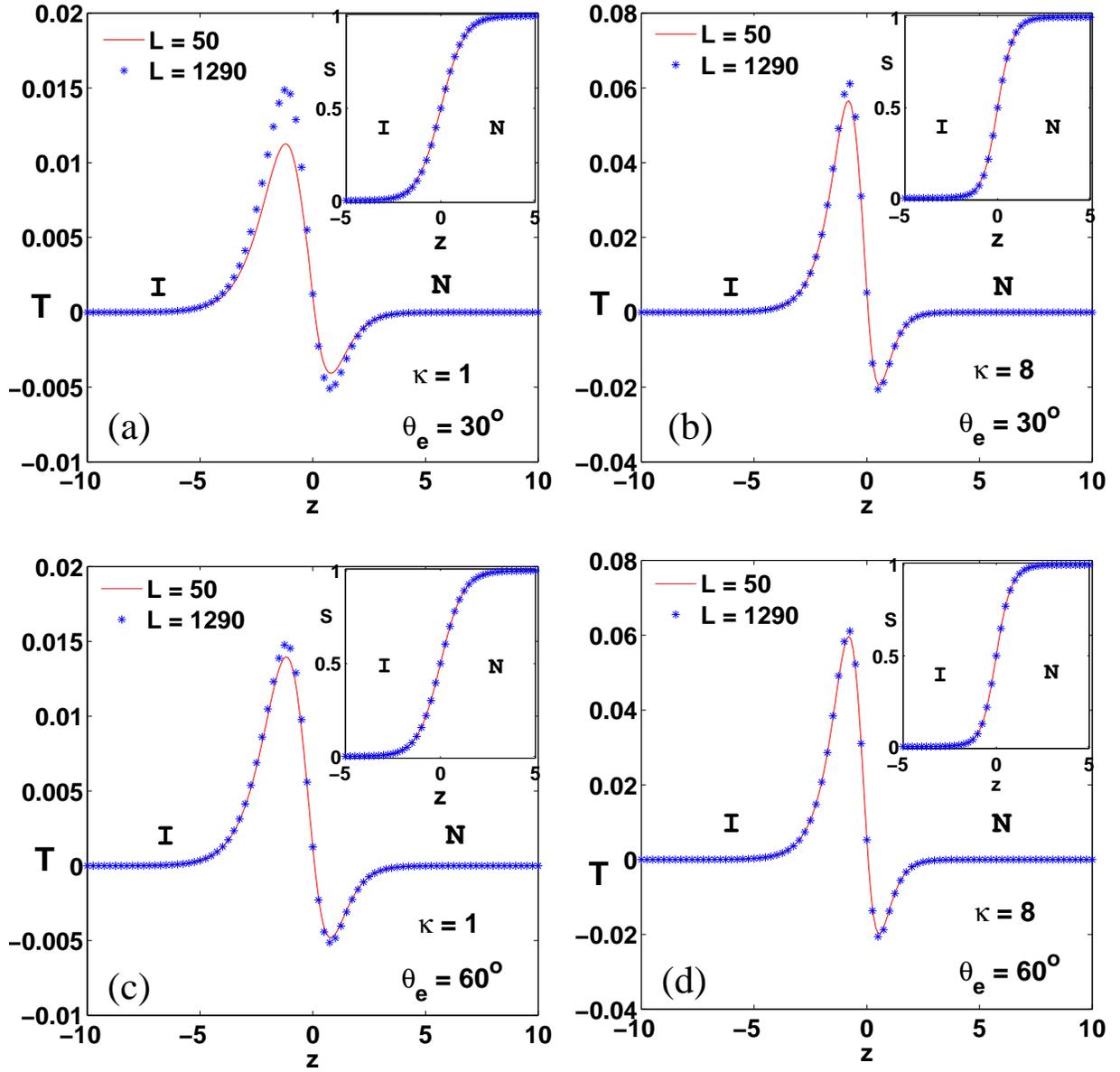}
\caption{ [Color online] Profiles of the biaxial ($T$) and uniaxial ($S$) order
parameter as a function of the coordinate $z$ across the interface, computed by
direct numerical minimization of the LGdG functional. These are shown in the main
figure for  systems of size $L=50, 1290$ and parameter values (a) $\kappa = 1, \theta_e = 30$
(b) $\kappa = 8, \theta_e = 30$, (c) $\kappa = 1, \theta_e = 60$
and (d) $\kappa = 8, \theta_e = 60$. The insets to each of (a), (b), (c) and (d) 
show the corresponding profiles for $S$. N and I refer to nematic and isotropic  
respectively.}
\label{jointcollapsT}
\end{figure}

\begin{figure}
\includegraphics[width = 7.5 in]{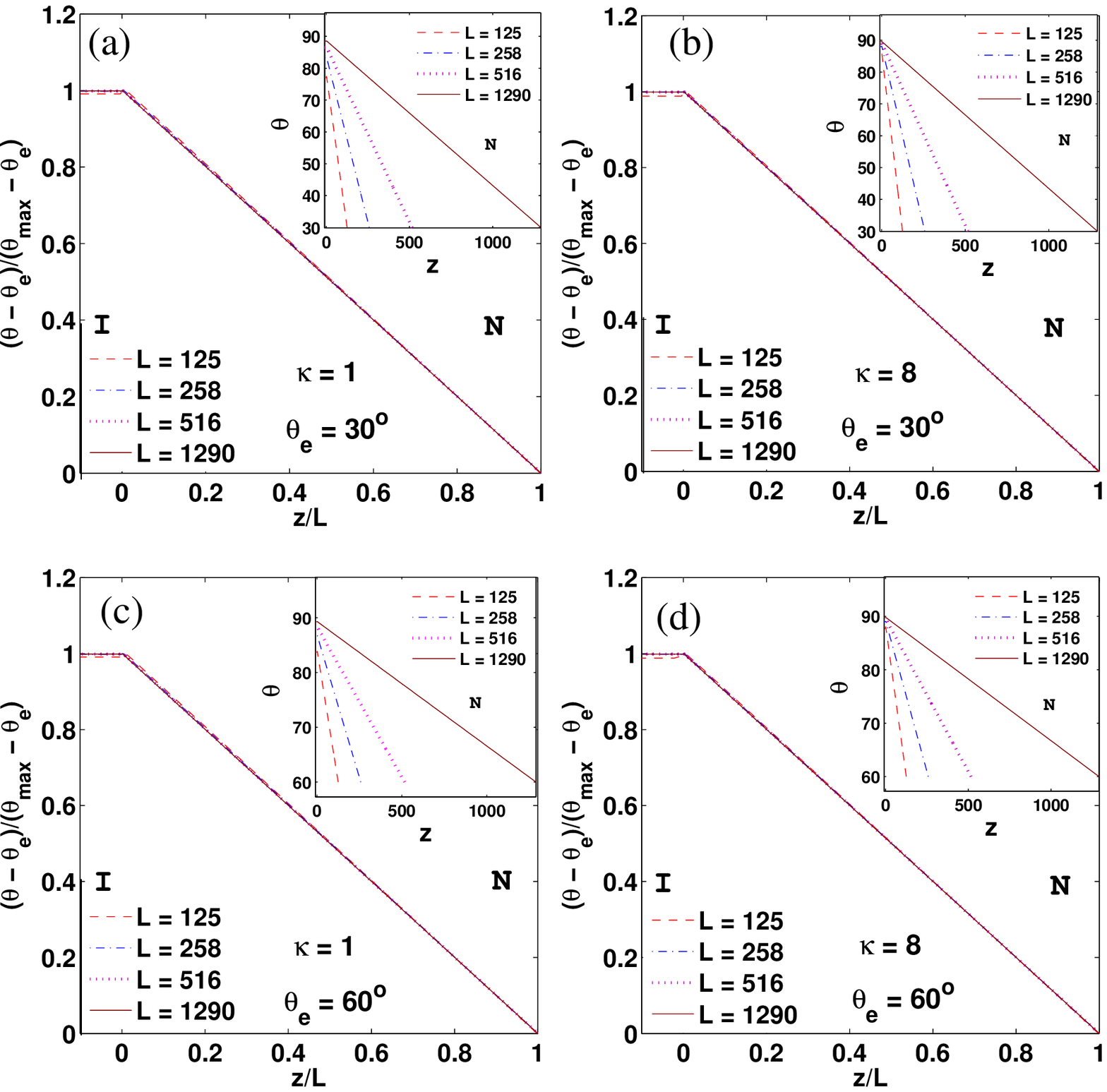}
\caption{[Color online] Main figure: Scaled profiles of the angle 
$\theta$ describing the orientation of the local director field as a function of the scaled
coordinate $z/L$ across the interface, as obtained from  direct numerical minimization of the LGdG functional
for systems of size  $L=125, 258, 516$ and $1290$. 
These are shown for  parameter values (a) $\kappa = 1, \theta_e = 30$
(b) $\kappa = 8, \theta_e = 30$, (c) $\kappa = 1, \theta_e = 60$
and (d) $\kappa = 8, \theta_e = 60$. The insets to each of (a), (b), (c) and (d) 
show the corresponding unscaled profiles for $\theta$.  }
\label{jointAngleprofile}
\end{figure}

We show the $T$ profile in the main sub-figure for systems of size $L=50, 1290$ and parameter values 
(a) $\kappa = 1, \theta_e = 30$ (b) $\kappa = 8, \theta_e = 30$, (c) $\kappa = 1, \theta_e = 60$
and (d) $\kappa = 8, \theta_e = 60$. N and I in the figure refer to nematic and isotropic  
respectively. The insets to each of (a), (b), (c) and (d)  show the corresponding profiles for $S$. 

We note that for larger anchoring angles, the $T$ profile converges faster as a function of system
size than for smaller angles; contrast the behavior for $\theta_e = 30^o$ and $\theta_e = 60^o$ 
in the figure. The profiles are qualitatively similar to profiles obtained for the $\theta_e = 90^o$ degree,
and asymptotically match this profile as $L \rightarrow \infty$. 

In the inset to Fig.~\ref{jointAngleprofile}, we show the profile of $\theta$, the angle 
describing the orientation of the local director field as a function of $z$ 
 across the interface, as obtained from  our  numerical minimization. We show data for 
systems of size  $L=125, 258, 516$ and $1290$, and for 
parameter values (a) $\kappa = 1, \theta_e = 30$
(b) $\kappa = 8, \theta_e = 30$, (c) $\kappa = 1, \theta_e = 60$
and (d) $\kappa = 8, \theta_e = 60$. The main figure, in each case, plots the same data as 
a function of the scaled coordinate $z/L$ on the $x-$axis  and the quantity 
$(\theta - \theta_e)/(\theta_{max} - \theta_e)$ on the $y-$axis , thus normalizing the
value to its maximum. This produces high quality collapse of the data, indicating
that the angle profile is linear on the nematic side, interpolating linearly between its
value at the interface to the anchored value of $\theta_e$ at $z = L$. Also, as the
system size is increased, the value at the interface ($z=0$), approaches $90^\circ$,
indicating that anchoring at the interface is always planar in the asymptotic limit.

In Fig.~\ref{jointcompNUMVAR} we show the comparison between the computed 
3-parameter variational profile for the angle 
$\theta$ as a function of the 
coordinate $z$ across the interface, for a system of size $L= 125$, as obtained from  a 
direct numerical minimization of the LGdG functional (solid line) and from
the variational calculation described in the text (point). These are shown for  parameter values 
(a) $\kappa = 1, \theta_e = 30$
(b) $\kappa = 8, \theta_e = 30$, (c) $\kappa = 1, \theta_e = 60$
and (d) $\kappa = 8, \theta_e = 60$.  The inset labeled (i) in each
sub-figure shows the corresponding profile of $S$, whereas the inset labeled (ii) shows
the profile of $T$. Note that the variational result coincides with the result obtained from
a direct numerical minimization to high accuracy. As the system size is increased, the value of
$\theta$ at the interface approaches $90^\circ$ within both the variational and the direct numerical
minimization approaches.

Fig.~\ref{varparamk1} shows  the variational parameters $a$ (a), $b$ (b)
and $c$ (c)  as a function of  system size $L$,  together with the variation of the variational angle 
$p$ (d), plotted for $\kappa = 1$. These parameters converge to their $L \rightarrow \infty$ values
corresponding to the case of planar anchoring. In all cases the parameter $p$ 
converges to the asymptotic value of $90^\circ$ as the system size is increased, consistent with
planar anchoring.

\begin{figure}
\includegraphics[width =7.5 in]{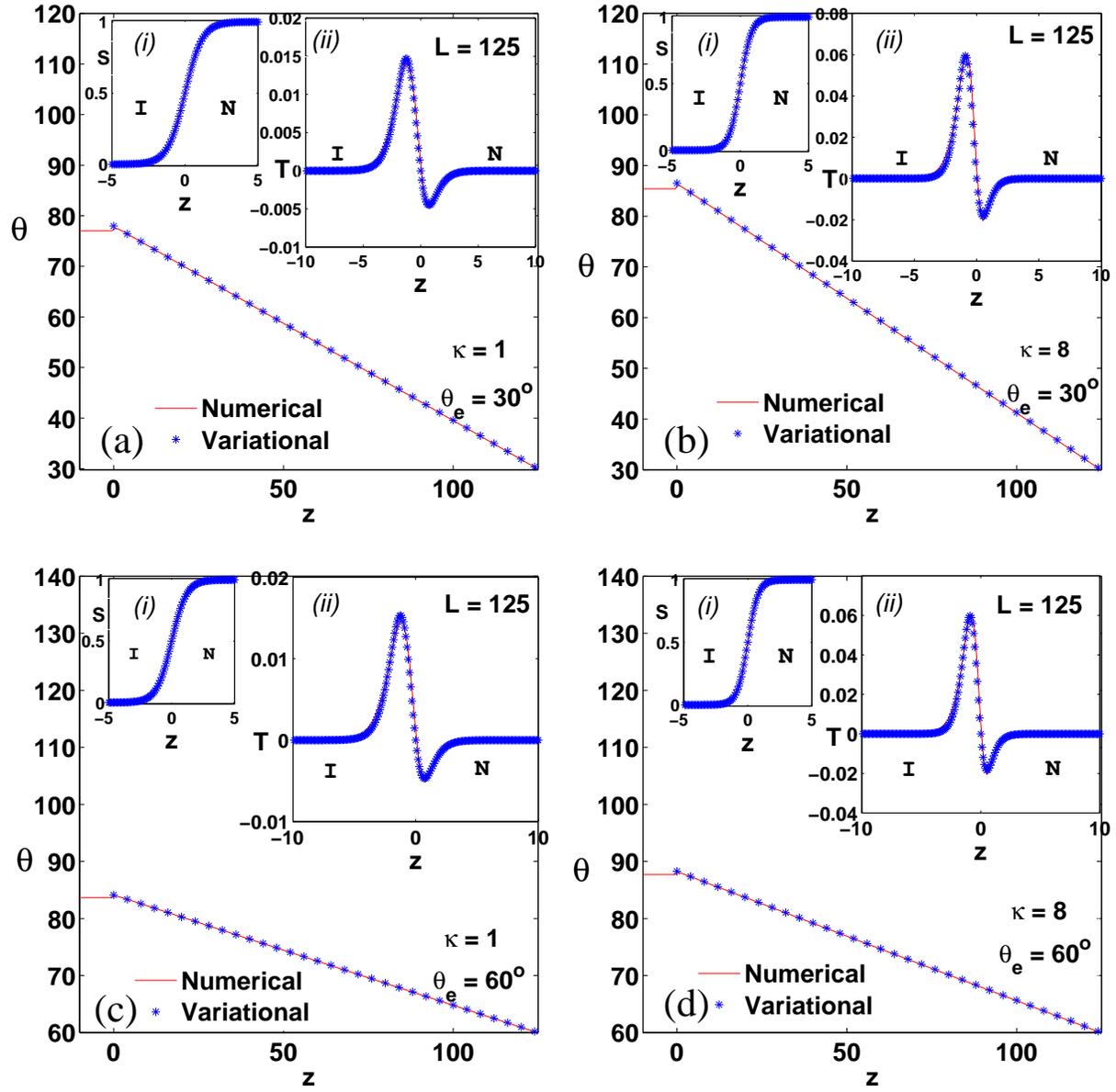}
\caption{[Color online] Main figure:  Profiles of the angle 
$\theta$ describing the orientation of the local director field as a function of the 
coordinate $z$ across the interface for a system of size $L= 125$, as obtained from  a 
direct numerical minimization of the LGdG functional (solid line) and from
the variational calculation described in the text (point). These are shown for  parameter values 
(a) $\kappa = 1, \theta_e = 30$
(b) $\kappa = 8, \theta_e = 30$, (c) $\kappa = 1, \theta_e = 60$
and (d) $\kappa = 8, \theta_e = 60$. 
The inset labeled (i) in each
sub-figure shows the corresponding profile of $S$, whereas the inset labeled (ii) shows
the profile of $T$.  N and I refer to nematic and isotropic,  respectively. }
\label{jointcompNUMVAR}
\end{figure}

\begin{figure}
\includegraphics[width = 7.5 in]{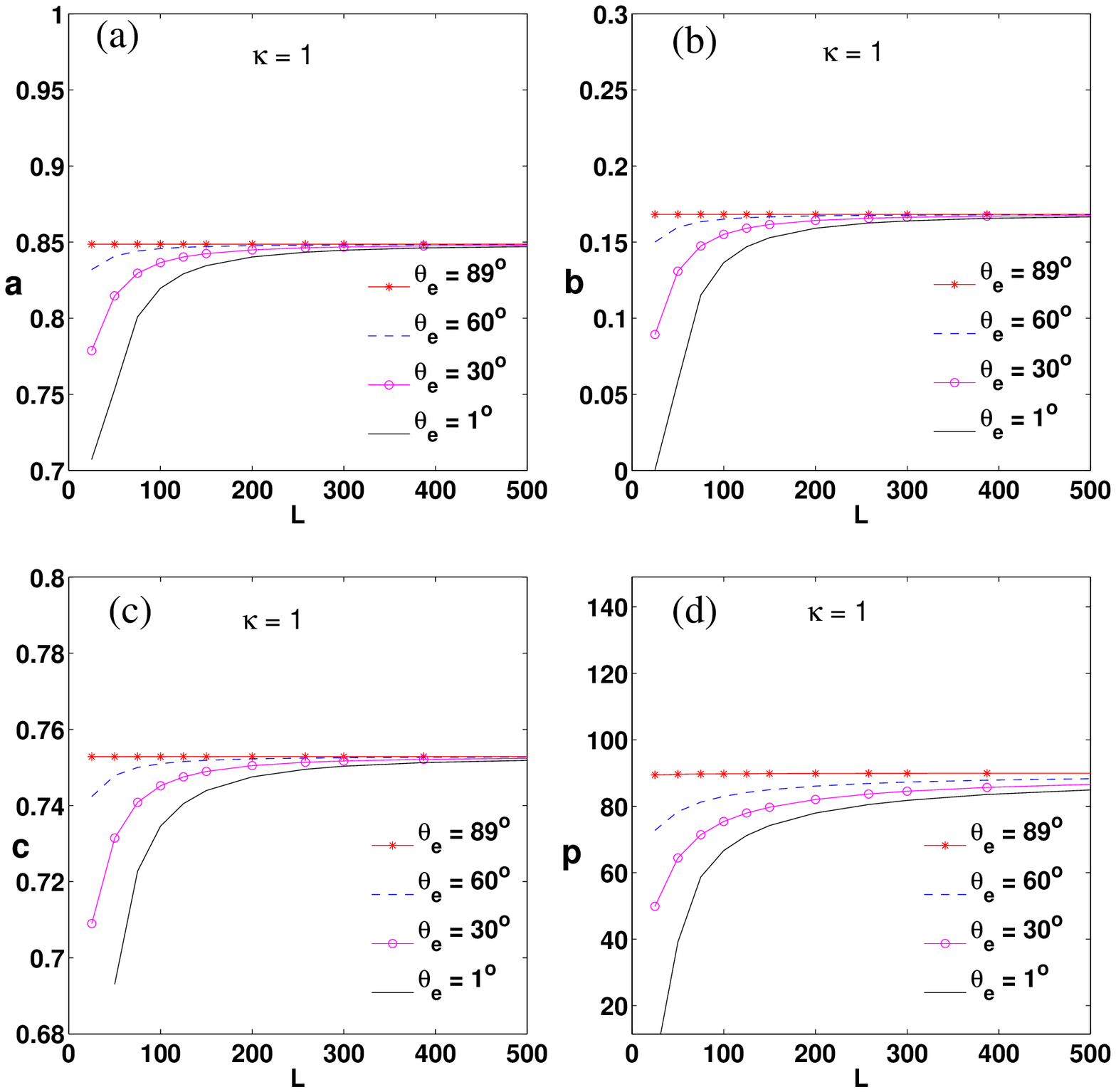}
\caption{[Color online] The variation of the  variational parameters $a$ (a), $b$ (b)
and $c$ (c) with system size $L$,  together with the variation of the variational angle 
$p$ (d), plotted for $\kappa = 1$. Note that these parameters quickly converge to their $L \rightarrow \infty$ values
corresponding to the case of planar anchoring. In all cases the parameter $p$ appears to
converge to the asymptotic value of $90^\circ$  as the system size is increased.
\label{varparamk1}
}
\end{figure}

%%%%%%%
\subsection{Results from the Numerical and Variational Minimization: $\kappa < 0$}
\label{sectionresultsnegkappa}

Stability imposes the requirement that $3 + 2\kappa > 0$, but does not constrain the
{\em sign} of $\kappa$ (or, equivalently $L_2$), apart from this requirement. In this
section we explore the consequences of a negative value for $L_2$. 
\begin{figure}
\includegraphics[width = 7.5 in]{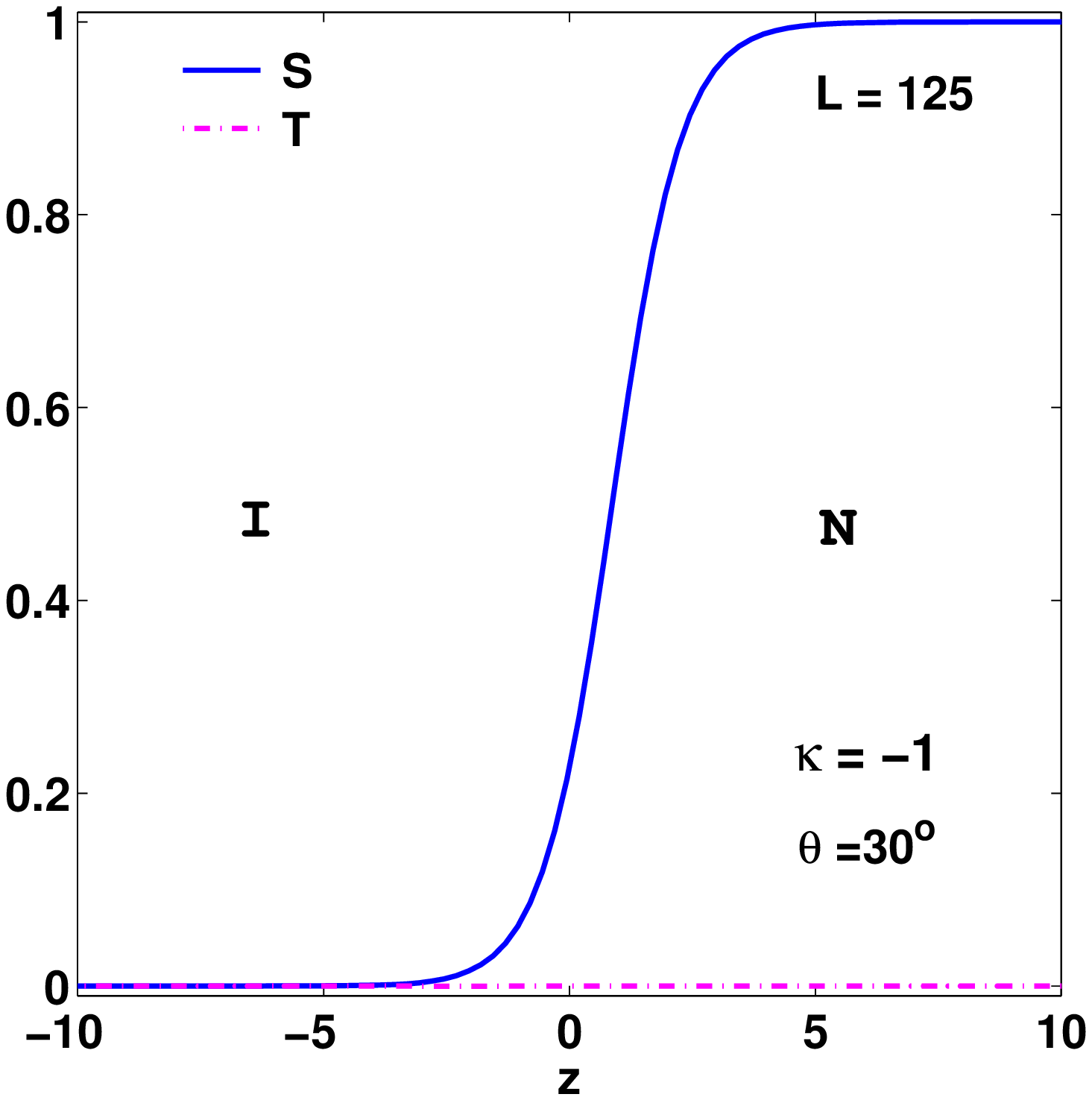}
\caption{[Color online]  The variation of the $S$ and $T$, for  system size $L = 125$,
plotted for  $\kappa = -1$, with an oblique anchoring angle of $30^\circ$. Our
results are consistent with $T=0$ for homoeotropic anchoring.
}
\label{STkminus1angle30}
\end{figure}

\begin{figure}
\includegraphics[width = 7.5 in]{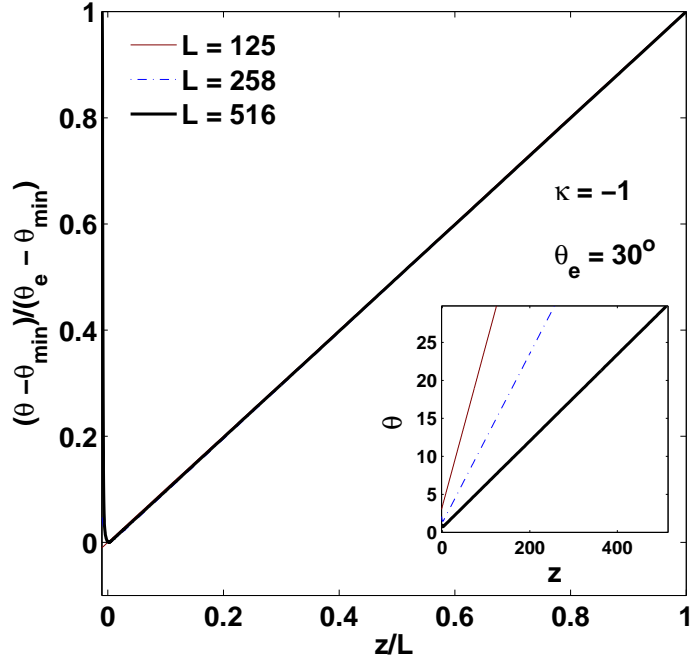}
\caption{[Color online]  Plot of the angle scaled to its minimum value for each system size ($L = 125,258$ and $516$,
against $z/L$ for $\kappa = -1$ and an asymptotic, oblique anchoring angle of $30^\circ$. The inset shows the bare angles
as a function of $z$ for these different system sizes. Note that the excellent 
data collapse indicates that angle profiles in the case of  $L_2 < 0$ scale in the same way as the $L_2 > 0 $ case, with
a homoeotropic anchoring being favoured at the interface.}
\label{scaledAngle30vsscaledDistancekminus1}
\end{figure}

\begin{figure}
\includegraphics[width = 6.5 in]{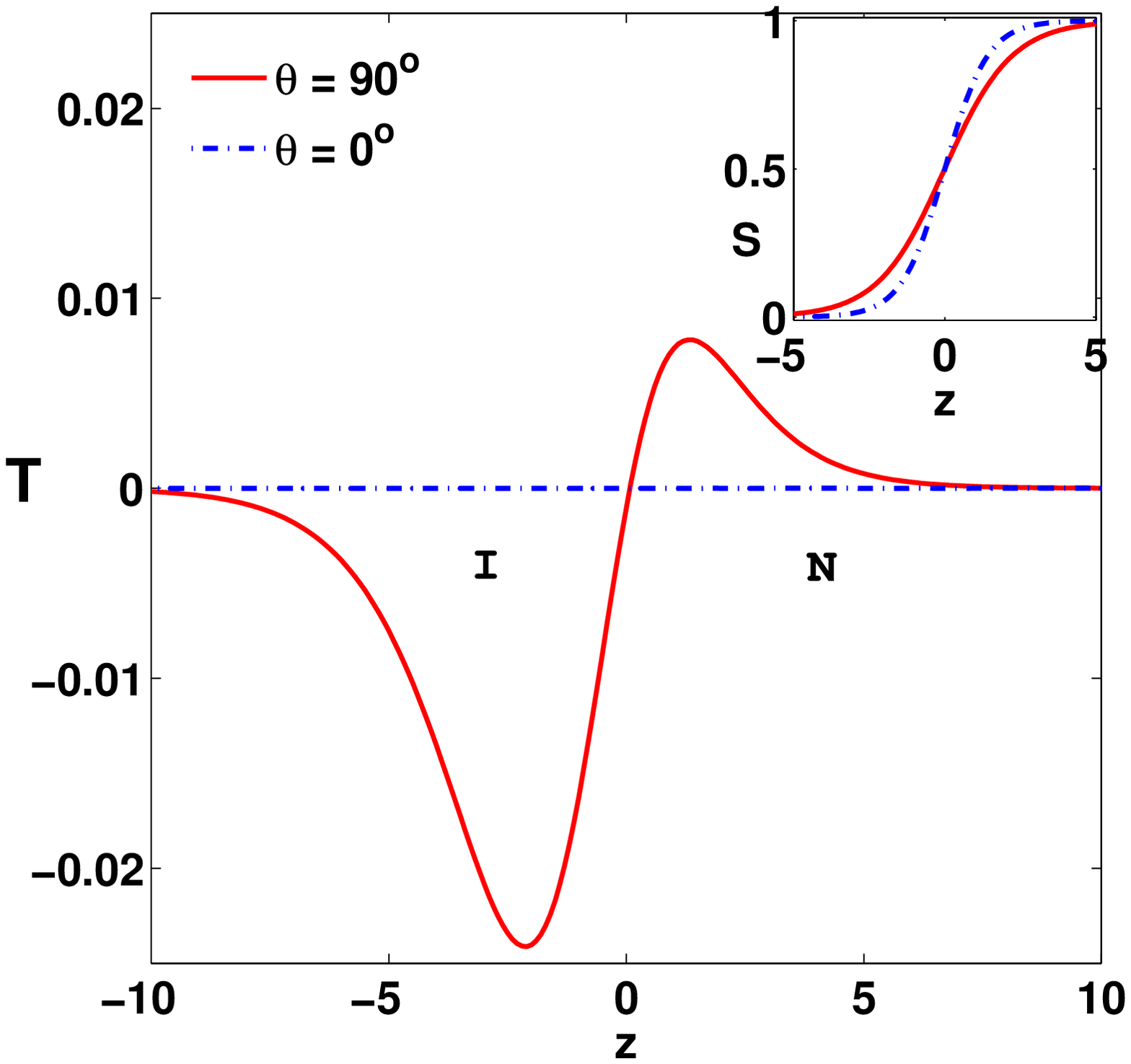}
\caption{[Color online]  Main Figure: Profile of $T$, the biaxial order parameter, for $\kappa = -1$,
in the two extreme cases of planar ($0^\circ$) and homoeotropic ($90^\circ$)
anchoring. Note that the profile of $T$ is {\em inverted}
with respect to profiles obtained for $\kappa > 0$, with the minimum appearing on the isotropic side
of the interface. INset: The profile of $S$, consistent with a tanh form. Data are computed for $L = 50$. While 
the profile of $T$ is non-zero for planar anchoring, biaxiality vanishes for the homoeotropic anchoring
case. }
\label{Tkminus1angle90and0}
\end{figure}

We find that, consistent with de Gennes' prediction,
a negative $L_2$ ( or $\kappa$) consistent with
stability favours {\em homoeotropic} anchoring at the interface, in contrast to
the case of positive $L_2$. Thus, the biaxiality $T$ generically vanishes as $L \rightarrow \infty$,
whereas $S$ assumes the canonical $\tanh$ form obtained by de Gennes. 
This can be seen from Fig.~\ref{STkminus1angle30} which shows the variation of  $S$ and $T$, 
for  $L = 125$, plotted for  $\kappa = -1$.
The anchoring at $L$ is set to   an oblique  angle of $30^\circ$. The $S$ and $T$ 
profiles  are consistent with $T=0$ for homoeotropic
anchoring.

The preference for homoeotropic anchoring can be seen from Fig.~\ref{scaledAngle30vsscaledDistancekminus1}
which shows the director tilt angle scaled to its minimum value for each system size ($L = 125,258$ and $516$,
against $z/L$ for $\kappa = -1$, where  an asymptotic, oblique anchoring angle of $30^\circ$ is imposed on the 
system at $L$. The inset shows the bare angles
as a function of $z$ for these different system sizes. The excellent 
data collapse indicates that angle profiles in the case of  $L_2 < 0$ scale in the same 
way as the $L_2 > 0 $ case, except that homoeotropic anchoring is favoured in this case.

Finally, in Fig.~\ref{Tkminus1angle90and0}, we show, in the main figure, the
profile of $T$, the biaxial order parameter, for $\kappa = -1$,
in the two extreme cases of planar ($0^\circ$) and homoeotropic ($90^\circ$)
anchoring, with $L=50$. Importantly, the profile of $T$ is {\em inverted}
with respect to profiles obtained for $\kappa > 0$, with the minimum appearing on the isotropic side
of the interface rather than the nematic side, as earlier. The profile of $S$ is consistent with a tanh form. 
While  the profile of $T$ is non-zero for planar anchoring, biaxiality vanishes for the homoeotropic
anchoring case.

These results are consistent with the general trends observed in the case of $\kappa  > 0$, with the
difference that homoeotropic, rather than planar, anchoring is preferred once $\kappa$ turns
negative.

%%%%%%%

\section{Asymptotic Solution}
\label{sectionasymp}
We can use our ansatz for $S$ and $T$ to check the self-consistency of our conjectured behaviour  for 
$\theta$. Our chosen forms imply $S = 1 -e^{-2az}$ and $T= -be^{-2az}$  deep into the nematic phase, as $
z \rightarrow \infty$.  Then $S'=2ae^{-2az}$, $T'=2abe^{-2az}$ and $S''= -4a^2e^{-2az}$, $T''=-4a^2be^{-2az}$.
Inserting these into the equation for $\theta$ as below,
\begin{eqnarray}
4 (2 + k)\left (3 S' + T' \right) \theta'- k \sin(2\theta) (S''- T''),  
+  2(2 + k) (3 S + T)\theta'' = 0,
\end{eqnarray}
we get
\begin{eqnarray}
8(2 + k)\left (3 +b \right)a e^{-2az} \theta'+  k \sin(2\theta) (1-b)a^2e^{-2az} 
+  6(2 + k)\theta'' = 0.
\end{eqnarray}
As $z\rightarrow \infty$, this equation  reduces to $\theta'' = 0$. Thus, $\theta$ should
have a linear profile in this asymptotic limit, taking the
form 
\be
\theta =  p + \psi \frac{z}{L}.
\ee

We can also compute corrections to this profile for $ z \rightarrow \infty^-$. Let us now expand
about the $z  = \infty$ limit, in which case $\theta'' =0$. Thus,
\begin{equation}
\frac{\theta'}{\sin(2\theta)} = \frac{-ka(1 - b)}{2(2+k)(3+b)}.
\end{equation} 
Integrating the left-hand side of this equation, we obtain
\be
\frac{1}{2} \ln \tan(\theta) - \ln C = \frac{-ka(1 - b)z}{2(2+k)(3+b)},
\ee
which has a solution
$\theta = \tan^{-1}[Ce^{\frac{-ka(1-b)z}{(2+k)(3+b)}}]$. It can be seen that this 
will vanish as z goes to $\infty$  and is, in effect, negligible apart from a region close to
the interface, at $z= 0$.

\section{Summary and Conclusions}
\label{sectionconclusion}

In this paper, we have presented our results for the problem of the isotropic-nematic
interface within Ginzburg-Landau-de Gennes theory, for the case in which
an oblique anchoring condition is imposed on the system asymptotically on the nematic
side. In this case, we find that
nematic elasticity dictates that  the nematic orientation smoothly interpolates between a value of 
$90^\circ$ at the interface (planar anchoring)  to the  anchored value at the boundary on the nematic side 
when $\kappa > 0$. Thus,  the preferred 
value of the anchoring angle at the interface is  $90^\circ$ in this case. The case $\kappa < 0$ with $\kappa$ satisfying the stability requirement
$\kappa > -1.5$ leads to stable homoeotropic anchoring at the interface, as predicted by de Gennes. 

We have used simple variationally based descriptions of the
structure of the interface, with our  methods capturing  essential features 
of interface structure, both qualitatively and  quantitatively, for the case of
oblique anchoring. Our methods
access the non-trivial structure of biaxiality at the interface, including the large tail 
towards the isotropic side and the change in the sign of the biaxial order parameter across the 
interface. Our approach also captures the inversion of the profile of biaxiality as $\kappa$ crosses zero.

The results presented here are broadly consistent with results from density functional approaches, molecular simulations 
and approaches based on the Onsager  functional, but necessitate fewer approximations, truncations or
assumptions about specific model systems. Thus, coarse-grained approaches based on the 
Ginzburg-Landau-de Gennes functional provide a powerful  methodology 
for understanding generic features of the isotropic-nematic interface.

\acknowledgements We thank C. Dasgupta and M. Muthukumar for useful discussions. 
This work was partially supported by the DST (India)
and the Indo-French Centre for the Promotion of Advanced Research.

\end{document}